\documentclass[12pt]{article}

\pdfoutput=1

\usepackage{graphicx}
\usepackage{hyperref}
\usepackage{amsmath}
\usepackage[table]{xcolor} 
\usepackage{color}
\begin{document}

\hoffset = -0.3truecm
\voffset = -1.1truecm

\title{\bf Half-Monopole in the Weinberg-Salam Model\footnote{Invited oral paper for ``The 8th Joint Meeting of Chinese Physicists Worldwide (OCPA8): Looking Forward to Quantum Frontiers
and Beyond International Conference on Physics and Education, Singapore (23 - 27 June 2014)"; To be submitted for publication}}

\author{
{\bf Rosy Teh\footnote{E-mail: rosyteh@usm.my}}, 
{\bf Ban-Loong Ng and Khai-Ming Wong}\\
{\normalsize School of Physics, Universiti Sains Malaysia}\\
{\normalsize 11800 USM Penang, Malaysia}}

\date{June 30, 2014}
\maketitle

\begin{abstract}
We present new axially symmetric half-monopole configuration of the SU(2)$\times$U(1) Weinberg-Salam model of electromagnetic and weak interactions. The half-monopole configuration possesses net magnetic charge $2\pi/e$ which is half the magnetic charge of a Cho-Maison monopole. The electromagnetic gauge potential is singular along the negative $z$-axis. However the total energy is finite and increases only logarithmically with increasing Higgs field self-coupling constant $\lambda^{1/2}$ at $\sin^2\theta_W=0.2312$. In the U(1) magnetic field, the half-monopole is just a one dimensional finite length line magnetic charge extending from the origin $r=0$ and lying along the negative $z$-axis. In the SU(2) 't Hooft magnetic field, it is a point magnetic charge located at $r=0$. The half-monopole possesses magnetic dipole moment that decreases exponentially fast with increasing Higgs field self-coupling constant $\lambda^{1/2}$ at $\sin^2\theta_W=0.2312$. 
\end{abstract}

\section{Introduction}

The monopole in the Maxwell theory was first discussed in 1931 by P.A.M. Dirac \cite{kn:1}. It is a point magnetic charge with a semi infinite string singularity and possesses infinite energy. It possesses magnetic charge $\frac{2\pi n}{e}$, where $e$ is the unit electric charge and $n$ an integer. Later in 1974, 't Hooft and Polyakov independently found the finite energy one monopole \cite{kn:2}. The 't Hooft-Polyakov monopole is a regular solution of the SU(2) Georgi-Glashow theory with pole strength $\frac{4\pi}{e}$ and a multimonopole with $n$ monopoles superimposed at one point possesses magnetic charge $\frac{4\pi n}{e}$ \cite{kn:3}. 

The Cho-Maison monopole of the SU(2)$\times$U(1) Weinberg-Salam theory was discussed in 1997 \cite{kn:4} and it is a hybrid between the Dirac monopole and the 't Hooft-Polyakov monopole. This monopole possesses infinite energy as the magnetic charge in the U(1) field is a point charge and hence the energy density blows up at the location of the monopole. However the mass of this monopole can be estimated \cite{kn:5} and is found to be within the range of the recent MoEDAL detector at LHC, CERN \cite{kn:6}. The magnetic charge of this hybrid monopole is $\frac{4\pi}{e}$. This electroweak monopole does not possess a string.

The Dirac, 't Hooft-Polyakov, and Cho-Maison monopoles possess radial symmetry and they are the only radially symmetrical solutions in their respective theories. All the other monopole configurations can at most possess axial symmetric. In the SU(2) Georgi-Glashow theory, other interesting monopole configurations include the single $n$-monopole \cite{kn:3}, the monopole-antimomopole pair (MAP), monopole-antimonopole chain (MAC), and the vortex-ring configurations \cite{kn:7}, \cite{kn:8} and they are all axially symmetric solutions. Recently axially symmetric finite energy half-monopole configurations are also found to exist in the SU(2) Georgi-Glashow theory. This half-monopole can exist by itself \cite{kn:9} or it can coexist together with a 't Hooft-Polyakov monopole \cite{kn:10}.  

Similarly in the SU(2)$\times$U(1) Weinberg-Salam theory, a rich variety of solutions has been found to exist \cite{kn:11}. An interesting solution is the sphaleron which was first coined by Klinkhamer and Manton \cite{kn:12} but predicted earlier by Y. Nambu \cite{kn:13}. He predicted the existence of massive string-like structure which is actually a monopole-antimonopole pair. This configuration is different from the MAP solutions of the the SU(2) Georgi-Glashow model in that the monopole-antimonopole pair is bound by a flux string of the ${\cal Z}^0$ field. The sphaleron possesses finite energy and magnetic dipole moment. The mass of the monopole and antimonopole together with the string is estimated to be in the TeV range. The sphaleron also possesses baryon number $Q_B=\frac{1}{2}$ and its' monopole-antimonopole pair is also surrounded by an electromagnetic current loop \cite{kn:12}, \cite{kn:14}, \cite{kn:15}. 

Other sphaleron configurations of the SU(2)$\times$U(1) Weinberg-Salam theory include the sphaleron-antisphaleron pairs, sphaleron-antisphaleron chains, and vortex-rings sphaleron \cite{kn:16}, \cite{kn:17}. These numerical solutions possess magnetic dipole moment and finite energy but failed to reveal the inner structure of the sphaleron and hence the source of the magnetic dipole moment in the solutions.

Recently, more monopoles and sphalerons solutions were found in the SU(2)$\times$U(1) Weinberg-Salam theory. These are the MAP, MAC, and vortex-ring configurations \cite{kn:18}. The MAP/vortex-ring configurations which possess zero net magnetic charge are actually the sphaleron (one monopole-antimonopole pair) and the sphaleron-antisphaleron pair (two monopole-antimonopole pairs), hence confirming the finding of others \cite{kn:11} - \cite{kn:15} that sphaleron does possess inner structure. The monopole and antimonopole in the sphaleron possess magnetic charges $\pm\frac{2\pi}{e}$ respectively and hence they are half Cho-Maison monopole and antimonopole and the Weinberg angle can only take the value $\theta_W=\frac{\pi}{4}$. The MAC/vortex-ring configurations that possess net magnetic charge $\frac{4\pi}{e}$ is a sequence of Cho-Maison monopole-antimomopole chains. The one Cho-Maison monopole \cite{kn:4} is the first member of this sequence of solutions.

In this paper, we present new monopole configuration that is axially symmetrical. The SU(2)$\times$U(1) Weinberg-Salam equations of motion are solved numerically for all space when the $\theta$-winding number $n=1$. This monopole configuration possesses magnetic charge $2\pi/e$ and hence it is a half Cho-Maison monopole. It possesses finite total energy even though the electromagnetic gauge potential is singular along the negative $z$-axis.  In the U(1) field, this half-monopole is just a one dimensional finite length line magnetic charge extending from the origin $r=0$ and lying along the negative $z$-axis.
The solution is studied by varying the Weinberg angle $\theta_W$ from $\frac{\pi}{18}$ rad to $\frac{\pi}{2}$ rad, when the Higgs field self-coupling constant $\lambda=1$, and also by varying the Higgs field self-coupling constant $\lambda$ when the Weinberg angle $\sin^2\theta_W=0.2312$. The Higgs field vacuum expectation value $\zeta$, and the unit electric charge is $e$ are both set to unity.
 
\section{The Standard Weinberg-Salam Model}
\label{sect.2}

Denoting the covariant derivative of the SU(2)$\times$U(1) group by ${\cal D}_\mu$ and the covariant derivative of the SU(2) group by $D_\mu$, the Lagrangian in the standard Weinberg-Salam model is written as \cite{kn:4} 
\begin{eqnarray}
&&{\cal L} = -{(\cal D}_\mu \boldsymbol{\phi})^\dagger ({\cal D}^\mu \boldsymbol{\phi}) - \frac{\lambda}{2}\left(\boldsymbol{\phi}^\dagger \boldsymbol{\phi} -\zeta^2\right)^2 - \frac{1}{4}{\bf F}_{\mu\nu}\cdot {\bf F}^{\mu\nu} - \frac{1}{4}G_{\mu\nu}G^{\mu\nu},
\label{eq.1}\\
&&{\cal D}_\mu \boldsymbol{\phi} = \left(D_\mu - \frac{ig^\prime}{2} B_\mu \right) \boldsymbol{\phi}, ~~D_\mu = \partial_\mu - \frac{ig}{2} \boldsymbol{\sigma} \cdot {\bf A}_\mu,
\label{eq.2}
\end{eqnarray}
where $\sigma^a$ are Pauli matrices and the metric used is $-g_{00}=g_{11}=g_{22}=g_{33}=1$.
The SU(2) gauge coupling constant, potentials, and electromagnetic fields are denoted by $g$, ${\bf A}_\mu = A^a_\mu (\frac{\sigma^a}{2i})$, and ${\bf F}_{\mu\nu} = F^a_{\mu\nu} (\frac{\sigma^a}{2i})$ respectively, whereas the U(1) gauge coupling constant, potentials, and electromagnetic fields are denoted by $g^\prime$, $B_\mu$, and $G_{\mu\nu}$ respectively.  The complex scalar Higgs doublet is $\phi$, the Higgs field self-coupling constant is $\lambda$, the Higgs field mass is $\mu$ and the Higgs field vacuum expectation value is given by $\zeta=\frac{\mu}{\sqrt{\lambda}}$. 

From Lagrangian (\ref{eq.1}), the equations of motion are found to be
\begin{eqnarray}
&&{\cal D^\mu}{\cal D_\mu}\boldsymbol{\phi} = \lambda\left(\boldsymbol{\phi}^\dagger\boldsymbol{\phi}-\zeta^2\right)\boldsymbol{\phi},
\label{eq.3}\\
&&D^\mu {\bf F}_{\mu\nu} = -{\bf j}_\nu = \frac{ig}{2}\{\boldsymbol{\phi}^\dagger\boldsymbol{\sigma}({\cal D_\nu}\boldsymbol{\phi})-({\cal D_\nu}\boldsymbol{\phi})^\dagger\boldsymbol{\sigma\phi}\},
\label{eq.4}\\
&&\partial^\mu G_{\mu\nu} = -k_\nu = \frac{ig^\prime}{2}\{\boldsymbol{\phi}^\dagger({\cal D_\nu}\boldsymbol{\phi})-({\cal D_\nu}\boldsymbol{\phi})^\dagger\boldsymbol{\phi}\}. 
\label{eq.5}
\end{eqnarray}

\noindent In order to simplify the equations of motion, the Higgs field is written as \cite{kn:4}
\begin{eqnarray}
\boldsymbol{\phi} = |\Phi|\boldsymbol{\xi}, ~~~\boldsymbol{\xi}^\dagger\boldsymbol{\xi}=1, ~~~
\hat{\Phi}^a = \boldsymbol{\xi}^\dagger\sigma^a \boldsymbol{\xi}, ~~~\sigma^a = \left(\begin{array}{ll}                   
																																												\delta^a_3 & \delta^a_1-i\delta^a_2\\
																																					 \delta^a_1+i\delta^a_2 & -\delta^a_3
																																					               \end{array}\right),
\label{eq.6}
\end{eqnarray}
where $|\Phi|$ is the Higgs modulus, $\boldsymbol{\xi}$ is a column 2-vector, and $\hat{\Phi}^a$ is the Higgs field unit vector. 

\section{The Axially Symmetric Magnetic Ansatz}
\label{sect.3}

The electrically neutral magnetic ansatz of the half-monopole configurations \cite{kn:9}, \cite{kn:10} is given by
\begin{eqnarray}
gA_i^a &=&  - \frac{1}{r}\psi_1(r, \theta) \hat{n}^{a}_\phi\hat{\theta}_i + \frac{1}{r\sin\theta}P_1(r, \theta)\hat{n}^{a}_\theta\hat{\phi}_i
+ \frac{1}{r}R_1(r, \theta)\hat{n}^{a}_\phi\hat{r}_i - \frac{1}{r\sin\theta}P_2(r, \theta)\hat{n}^{a}_r\hat{\phi}_i, \nonumber\\
gA^a_0 &=& 0, ~g\Phi^a = \Phi_1(r, \theta)\hat{n}^a_r + \Phi_2(r, \theta)\hat{n}^a_\theta = \Phi(r, \theta) \hat{\Phi}^a, 
\label{eq.7}\\
g^\prime B_i &=& \frac{1}{r\sin\theta}{\cal B}_S(r,\theta)\hat{\phi}_i, ~g^\prime B_0=0, ~\boldsymbol{\xi} = i\left(\begin{array}{l}                   
																																												\sin\frac{\alpha(r,\theta)}{2}e^{-in\phi}\\
																																					 							 -\cos\frac{\alpha(r,\theta)}{2}
																																					               \end{array}\right), ~
\hat{\Phi}^a = \boldsymbol{\xi}^\dagger\sigma^a \boldsymbol{\xi} = -\hat{h}^a, \nonumber																																				      
\end{eqnarray}
where the Higgs modulus, $g|\Phi|= \Phi = \sqrt{\Phi_1^2 + \Phi_2^2}$ and the unit vector, \cite{kn:19}
\begin{eqnarray}
\hat{h}^a&=&h_1\hat{n}^{a}_r + h_2\hat{n}^{a}_\theta = \sin\alpha \cos n\phi ~\delta^{a1} + \sin\alpha \sin n\phi ~\delta^{a2} + \cos\alpha ~\delta^{a3},  
\label{eq.8}\\
h_1 &=&\cos(\alpha-\theta), ~~h_2 =\sin(\alpha-\theta), ~~\alpha=\alpha(r,\theta). \nonumber
\end{eqnarray}

\noindent In the half-monopole solutions of the SU(2) Georgi-Glashow model, the angle $\alpha(r, \theta) \rightarrow \frac{1}{2}\theta$ as $r\rightarrow \infty$ \cite{kn:9}, \cite{kn:10}. 
The isospin coordinate unit vectors with $\phi$-winding number $n=1$ are given by
\begin{eqnarray}
\hat{n}_r^a&=&\sin \theta ~\cos n\phi ~\delta_{1}^a + \sin \theta ~\sin n\phi ~\delta_{2}^a + \cos \theta~\delta_{3}^a, ~\hat{n}_\phi^a = -\sin n\phi ~\delta_{1}^a + \cos n\phi ~\delta_{2}^a\nonumber\\
\hat{n}_\theta^a&=&\cos \theta ~\cos n\phi ~\delta_{1}^a + \cos \theta ~\sin n\phi ~\delta_{2}^a - \sin \theta ~\delta_{3}^a. 
\label{eq.9}
\end{eqnarray}

The magnetic ansatz (\ref{eq.7}) is substituted into the equations of motion (\ref{eq.3}) to (\ref{eq.5}) and the total number of equations of motions is reduced to only seven second order nonlinear coupled partial differential equations \cite{kn:18}.

In the electrically neutral monopole configuration, the energy density can be written as 
\begin{eqnarray}
e^2{\cal E}_n &=& \cos^2\theta_W ~{\cal E}_0 + \sin^2\theta_W ~{\cal E}_1 + {\cal E}_H, \nonumber\\
{\cal E}_0 &=& \frac{g^{\prime 2}}{4} G_{ij} G_{ij}, ~~~{\cal E}_1 = \frac{g^2}{4}F^a_{ij}F^a_{ij}\nonumber\\
{\cal E}_H &=& \sin^2\theta_W\partial^i\Phi\partial_i\Phi + \sin^2\theta_W\Phi^2({\cal D}^i\xi)^\dagger({\cal D}_i\xi) + \frac{\lambda}{2}\left(\sin^2\theta_W\Phi^2 -\zeta^2\right)^2,
\label{eq.10}
\end{eqnarray}
\begin{eqnarray}
({\cal D}^i\xi)^\dagger({\cal D}_i\xi)&=&\frac{1}{4}\partial^i\alpha\,\partial_i\alpha + \frac{n^2(1-\cos\alpha)}{2r^2\sin^2\theta} + \frac{n}{2}(1-\cos\alpha)(g^\prime B^i)\partial_i\phi\nonumber\\
&+&\frac{1}{2}\{\hat{n}_\phi^a \partial^i\alpha + n\,\partial^i\phi\,[\hat{n}_r^a\cos\theta - \hat{n}_\theta^a\sin\theta - \hat{h}^a]\}(gA^a_i)\nonumber\\
&+&\frac{1}{4}(gA^{ai})(gA^a_i)-\frac{1}{2}(g^\prime B^i)(gA^a_i)\hat{h}^a + \frac{1}{4}(g^\prime B^i)(g^\prime B_i).
\label{eq.11}
\end{eqnarray}
\noindent The total energy is given by $E = \frac{e}{4\pi}\int{{\cal E}_n}\,d^3x$.

We choose to define the electromagnetic gauge potential and the neutral ${\cal Z}^0$ gauge potential by first gauge transforming the gauge potentials $A^a_\mu$ and Higgs field $\Phi^a$ of Eq. (\ref{eq.7}) to $A^{\prime a}_\mu$ and $\Phi^{\prime a}=\delta^a_3$ using the gauge transformation, \cite{kn:4} 

\begin{eqnarray}
&&U = -i\left[\begin{array}{ll}                   
																 \cos\frac{\alpha}{2}  				  & \sin\frac{\alpha}{2} e^{-in\phi}\\
																 \sin\frac{\alpha}{2} e^{in\phi} & -\cos\frac{\alpha}{2}
																 \end{array}\right]
= \cos\frac{\Theta}{2} + i\hat{u}_r^a \sigma^a \sin\frac{\Theta}{2}, 														 
\label{eq.12}\\
&&\Theta=-\pi ~~\mbox{and}~ ~\hat{u}_r^a = \sin\frac{\alpha}{2}\cos n\phi \delta^a_1 + \sin\frac{\alpha}{2}\sin n\phi \delta^a_2 + \cos\frac{\alpha}{2} \delta^a_3.	\nonumber	
\end{eqnarray}																
																 
\noindent The transformed Higgs column unit vector and the SU(2) gauge potentials in the unitary gauge are 
\begin{eqnarray}
\xi^\prime &=& U\xi = \left[\begin{array}{l}                   
																 0\\
																 1
																 \end{array}\right]\nonumber\\
gA^{\prime a}_\mu &=& -gA^a_\mu - \frac{2}{r}\left\{\psi_2\sin\left(\theta-\frac{\alpha}{2}\right) + R_2\cos\left(\theta-\frac{\alpha}{2}\right)\right\}\hat{u}_r^a\,\hat{\phi}_\mu\nonumber\\
 &-& \partial_\mu\alpha\,\hat{u}_\phi^a - \frac{2n\sin\frac{\alpha}{2}}{r\sin\theta}\hat{u}_\theta^a\,\hat{\phi}_\mu.															 
\label{eq.13}
\end{eqnarray}																

\noindent Subsequently the electromagnetic gauge potential ${\cal A}_\mu$ and the neutral gauge potential ${\cal Z}_\mu$ are defined as
\begin{eqnarray}
\left[\begin{array}{l}                   
			{\cal A}_\mu\\
			{\cal Z}_\mu
			\end{array}\right] = \left[\begin{array}{ll} 
			                      \cos\theta_W & \sin\theta_W \\
			                     -\sin\theta_W &  \cos\theta_W 
			                      \end{array}\right]         
\left[\begin{array}{l} 
B_\mu \\
A^{\prime 3}_\mu 
\end{array}\right]\nonumber\\\nonumber\\
= \frac{1}{\sqrt{g^2+g^{\prime 2}}}	\left[\begin{array}{ll} 
			                             g & g^\prime \\
			                     -g^\prime &  g
			                      \end{array}\right]         
\left[\begin{array}{l} 
B_\mu \\
A^{\prime 3}_\mu 
\end{array}\right]	                     			                   															 
\label{eq.14}
\end{eqnarray}	
where 
\begin{eqnarray}
gA^{\prime 3}_\mu = \frac{1}{r}\left\{\psi_2 h_2 - R_2 h_1 - \frac{n(1-\cos\alpha)}{\sin\theta}\right\}\hat{\phi}_\mu
\label{eq.15}
\end{eqnarray}
is recognized to be the 't Hooft gauge potential \cite{kn:18}. The Weinberg angle is $\theta_W=\cos^{-1}\frac{g}{\sqrt{g^2+g^{\prime 2}}}$.

The mass of $W^\pm$, $Z^0$, and Higgs bosons are given respectively by
\begin{eqnarray}
M_W = \frac{g\zeta}{\sqrt{2}}, ~~M_Z = \frac{\zeta\sqrt{g^2+g^{\prime 2}}}{\sqrt{2}}, ~~\mbox{and}~~ M_H = \sqrt{2}\mu.
\label{eq.16}
\end{eqnarray}
Hence $\frac{M_W}{M_Z}=\cos\theta_W$ and by using the experimental values for the mass of the $W^\pm$ and $Z^0$ bosons, where $M_W = 80.385(15)$ GeV and $M_Z = 91.1876(21)$ GeV \cite{kn:20}, the Weinberg angle can be calculated to be $\theta_W = 28.74^o$ ($\sin^2\theta_W=0.2312$), although in the standard model, the angle $\theta_W$ is an arbitrary parameter.

\section{The Half-Monopole Configuration}
\label{sect.4}

\subsection{Numerical Procedure}
\label{sect.4.1}

Using the Maple and MATLAB software, the Weinberg-Salam equations of the motions were solved numerically for all space by solving for the profiles functions, $\psi_1$, $P_1$, $R_1$, $P_2$, $\Phi_1$, $\Phi_2$, and ${\cal B}_S$ . The seven reduced coupled second order partial differential equations are solved by fixing boundary conditions at small distances ($r\rightarrow 0$), large distances ($r\rightarrow \infty$), and along the $z$-axis at $\theta=0$ and $\pi$ of the seven profile functions \cite{kn:9}, \cite{kn:10}, \cite{kn:18}.

The asymptotic solutions at large $r$ are the self-dual solution \cite{kn:9}, \cite{kn:10}
\begin{eqnarray}
&&\psi_1 = \frac{1}{2}, ~~P_1 =\sin\theta - \frac{1}{2}\sin \left(\frac{1}{2}\theta\right) (1+\cos\theta),\nonumber\\
&&R_1 =0, ~~P_2 = \cos\theta - \frac{1}{2}\cos \left(\frac{1}{2}\theta\right) (1+\cos\theta)
\label{eq.17}\\
&&\Phi_1 = \zeta \cos\left(\frac{1}{2}\theta\right), ~~\Phi_2 = -\zeta \sin\left(\frac{1}{2}\theta\right), ~~{\cal B}_S = - \frac{1}{2}(1-\cos\theta).\nonumber
\end{eqnarray}
The asymptotic solution at small $r$ is the trivial vacuum solution,
\begin{eqnarray}
\psi_1(0, \theta) = P_1(0, \theta) = R_1(0, \theta) = P_2(0, \theta) = {\cal B}_S(0, \theta) &=& 0,\nonumber\\
\sin\theta ~\Phi_1(0,\theta) + \cos\theta ~\Phi_2(0,\theta) &=& 0,\nonumber\\
\left.\frac{\partial}{\partial r}\left\{\cos\theta ~\Phi_1(r,\theta) - \sin\theta ~\Phi_2(r,\theta)\right\}\right|_{r=0} &=& 0.
\label{eq.18}
\end{eqnarray}

\noindent The common boundary condition of the profile functions along the positive $z$-axis at $\theta=0$ is
\begin{eqnarray}
\partial_\theta \psi_1 = R_1 = P_1 = P_2 = \partial_\theta \Phi_1 = \Phi_2 ={\cal B}_S = 0.
\label{eq.19}
\end{eqnarray}
Along the negative $z$-axis, the boundary condition imposed upon the profile functions is
\begin{eqnarray}
\partial_\theta \psi_1 = R_1 = P_1 = \partial_\theta P_2 = \Phi_1 = \partial_\theta \Phi_2 = \partial_\theta {\cal B}_S = 0.
\label{eq.20}
\end{eqnarray}

From Eq. (\ref{eq.14}), the electromagnetic gauge potential and the neutral ${\cal Z}^0$ field gauge potential can also be written as
\begin{eqnarray}
{\cal A}_\mu &=& \frac{1}{e} \left(\cos^2\theta_W g^\prime B_\mu + \sin^2\theta_W g A^{\prime 3}_\mu\right)\nonumber\\
{\cal Z}_\mu &=& \frac{1}{e}\cos\theta_W \sin\theta_W\left(-g^\prime B_\mu + g A^{\prime 3}_\mu\right),
\label{eq.21}
\end{eqnarray}
where the unit electric charge $e=\frac{gg^\prime}{\sqrt{g^2+g^{\prime 2}}}$.
For the monopole solutions presented here, the boundary conditions at large $r$ (\ref{eq.17}) is such that $g A^{\prime 3}_\mu \rightarrow g^\prime B_\mu$ and the neutral gauge potential ${\cal Z}_\mu$ vanishes at large distances. Hence this neutral ${\cal Z}^0$ field carries zero net electric and magnetic charges as expected. The electromagnetic gauge potential 
${\cal A}_\mu \rightarrow \frac{1}{e}(g^\prime B_\mu)$ at spatial infinity and the boundary condition for the half-monopole solution is such that $g^\prime B_i = -\frac{(1-\cos\theta)}{2r\sin\theta}\hat{\phi}_i$ at large $r$. Hence the half-monopole solution possesses magnetic charge $\frac{2\pi}{e}$. 
The electromagnetic dipole moment $\mu_m$ can also be calculated by using the boundary condition of the electromagnetic gauge potential at large $r$,
\begin{eqnarray}
{\cal A}_i \rightarrow \frac{1}{e}(g^\prime B_i) = \frac{1}{e}{\cal B}_S\,\partial_i\phi = -\frac{\hat{\phi}_i}{r\sin\theta}\left(\frac{\mu_m\sin^2\theta}{r}\right).
\label{eq.22}
\end{eqnarray}
Hence $r{\cal B}_S=-e\mu_m\sin^2\theta$ and by plotting the numerical result for $r{\cal B}_S$, we can read the magnetic dipole moment in unit of $\frac{1}{e}$ at $\theta=\frac{\pi}{2}$.

As in Ref. \cite{kn:18}, the seven reduced equations of motion were converted into a system of nonlinear equations using the finite difference approximation method, which is then discretized onto a non-equidistant grid of size 70 $\times$ 60 covering the integration regions $0\leq \bar{x} \leq 1$ and $0\leq \theta\leq \pi$. The compactified coordinate $\bar{x} = \frac{r}{r+1}$ runs from zero to unity. Upon replacing the partial derivative ~$\partial_r \rightarrow (1-\bar{x})^2 \partial_{\bar{x}}$~ and ~$\frac{\partial^2}{\partial r^2} \rightarrow (1-\bar{x})^4\frac{\partial^2}{\partial \bar{x}^2} - 2(1-\bar{x})^3\frac{\partial}{\partial \bar{x}}$~, the Jacobian sparsity pattern of the system was constructed by using Maple. The system of nonlinear equations is then solved numerically by MATLAB using the constructed Jacobian sparsity pattern, the trust-region-reflective algorithm, and a good initial starting solution. The overall error in the numerical results is estimated at $10^{-4}$.

\subsection{Half-Monopole Configuration}
\label{sect.4.2}

The half-monopole solution is solved numerically by setting the unit electric charge and the Higgs field vacuum expectation value to unity, that is $e=\zeta=1$. The solution is studied by setting the Higgs field self-coupling constant $\lambda = 1$ and then varying the Weinberg angle $\theta_W$ from $\frac{\pi}{18}$ rad to $\frac{\pi}{2}$ rad. Using the experimental value of the Weinberg angle, $\sin^2\theta_W=0.2312$, the solution is then solved for various values of $\lambda$ from zero to 40. 

\begin{figure}[tbh]
\centering
\hskip-0.1in
\includegraphics[width=6.0in,height=6.0in]{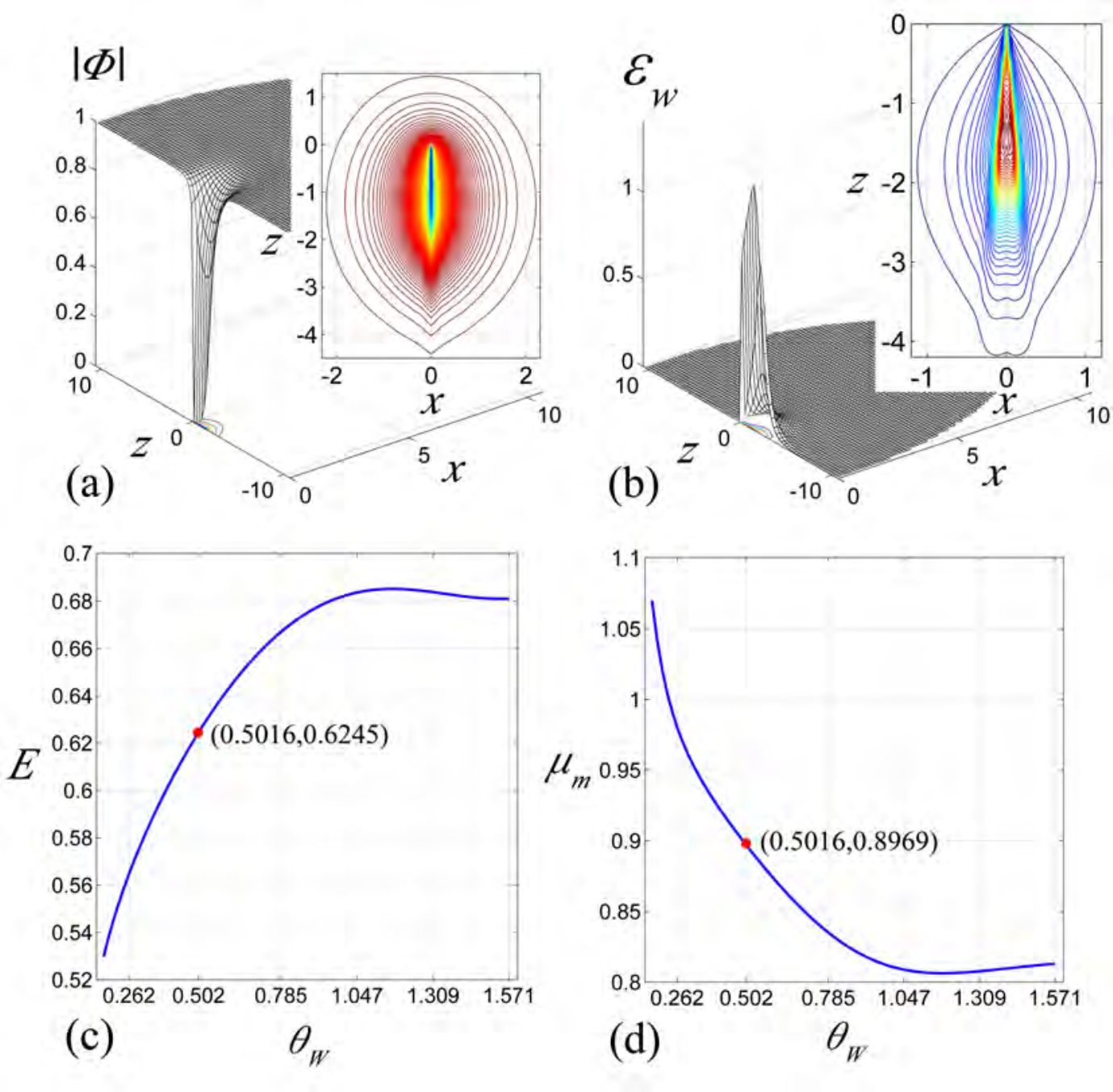} 
\caption{3D and contour line plots of (a) the Higgs field modulus $|\Phi|$ and (b) the weighted energy density ${\cal E}_W$ along the $x$-$z$ plane when $\sin^2\theta_W=0.2312$ and $\lambda=1$. The plots of (c) total energy $E$ in units of $\frac{1}{4\pi}$ and (d) magnetic dipole moment $\mu_m$ versus the Weinberg angle $\theta_W$ in radians when $\lambda=1$.}             
\label{fig.1}
\end{figure}

The numerical result is all the seven profile functions, $\psi_1$, $P_1$, $R_1$, $P_2$, $\Phi_1$, $\Phi_2$, and ${\cal B}_S$ are smooth regular bounded functions of $r$ and $\theta$. However, $P_2(r,\theta)|_{\theta=\pi}$ and ${\cal B}_S(r,\theta)|_{\theta=\pi}$ are both nonzero along the negative $z$-axis and they vary from zero at $r=0$ to negative one at $r=\infty$ along the $z$-axis. Hence the SU(2) and U(1) gauge potentials are singular along the negative $z$-axis. The 3D and contour line plots of the Higgs modulus $|\Phi|$ of the half-monopole solution are shown in Figure \ref{fig.1} (a) for $\zeta=\lambda=1$ and $\sin^2\theta_W=0.2312$. The shape and size of the graphs are almost similar to that of the SU(2) Georgi-Glashow half-monopole \cite{kn:9}. 

The semi-infinite line singularity of the SU(2) and U(1) gauge potentials along the negative $z$-axis is integrable and hence the weighted energy density ${\cal E}_W = 2\pi r^2\sin\theta\,{\cal E}$ does not blow up along the negative $z$-axis. By taking the Weinberg angle $\theta_W=28.74^o$, the 3D and contour line plots of weighted energy density ${\cal E}_W$ are shown in Figure \ref{fig.1} (b) for $\zeta=\lambda=1$. The energy of the half-monopole is concentrated along a finite length of the negative $z$-axis extending from the origin at $r=0$. 

The numerical values of the total energy $E$ and the magnetic dipole moment $\mu_m$ are tabulated in Table \ref{table.1} for values of $\frac{\pi}{18}\leq\theta_W\leq \frac{\pi}{2}$ when $\lambda=1$ and in Table \ref{table.2} for values of $0<\lambda\leq 40$ when $\sin^2\theta_W=0.2312$.  The plots of energy $E$ and magnetic dipole moment $\mu_m$ versus Weinberg angle $\theta_W$ when $\lambda=1$ are shown in Figure \ref{fig.1} (c) and (d) respectively.
The energy of the half-monopole here increases logarithmically with increasing $\theta_W$ until $\theta_W\approx 1.169$ rad ($67^o$) when $E=0.6852$ and decreases to $E=0.6811$ at $\theta_W=\frac{\pi}{2}$. At the experimental value of $\theta_W=0.5016$ rad, the energy $E=0.6245$. Similarly the magnetic dipole moment possess a turning point at the same angle $\theta_W\approx 1.169$ rad. It decreases exponentially with $\theta_W$ until $\theta_W\approx 1.169$ rad when $\mu_m= 0.8064$ and then increases to $\mu_m=0.8127$ at  $\theta_W=\frac{\pi}{2}$. At the experimental value of $\theta_W=0.5016$ rad, the magnetic dipole moment $\mu_m=0.8969$.

\begin{figure}[tbh]
\centering
\hskip-0.1in
\includegraphics[width=6.0in,height=6.0in]{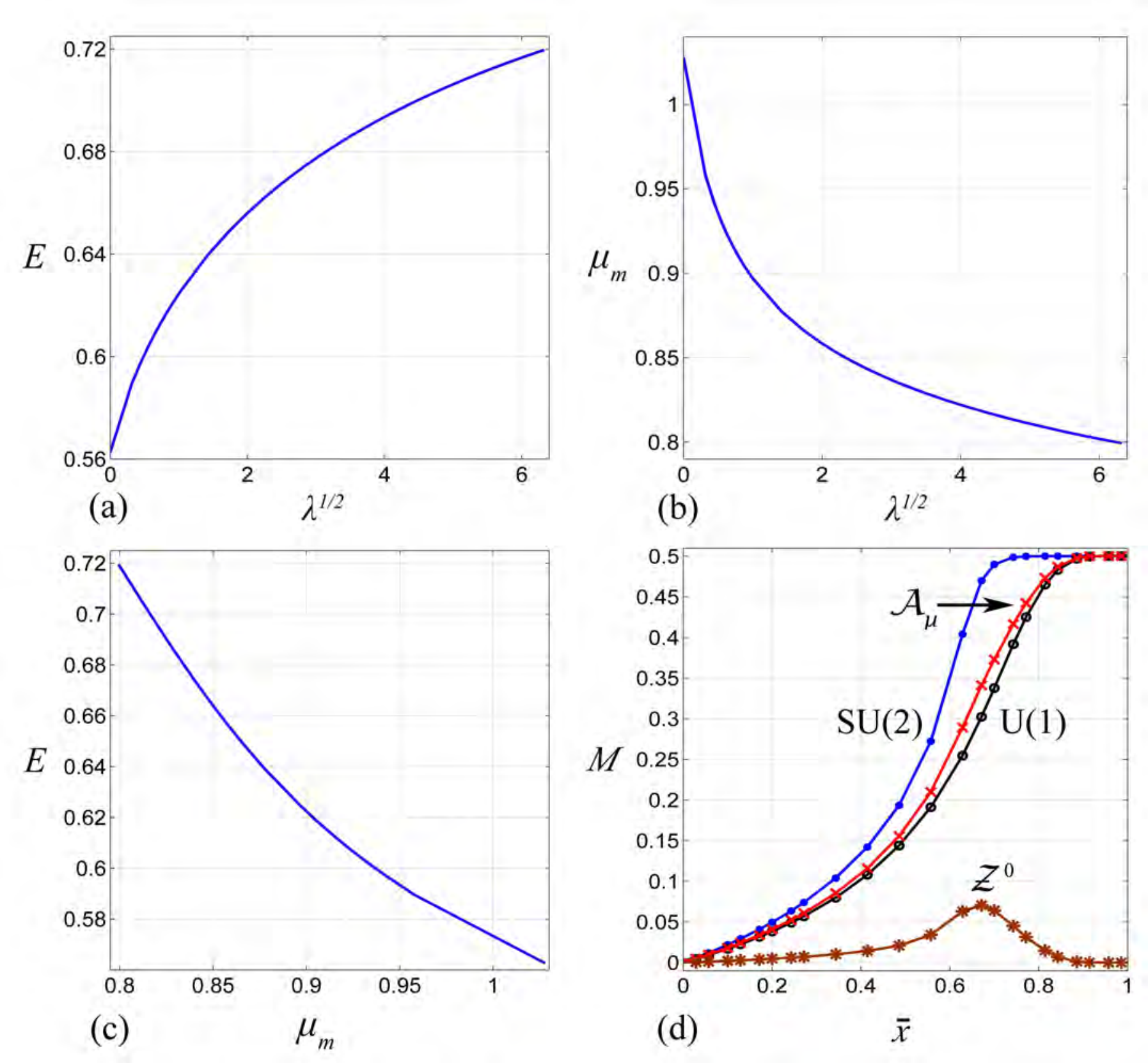} 
\caption{The plots of (a) total energy $E$ in units of $\frac{1}{4\pi}$ and (b) magnetic dipole moment $\mu_m$ versus the $\lambda^{1/2}$ when $\sin^2\theta_W=0.2312$. (c) The plot of total energy $E$ in units of $\frac{1}{4\pi}$ versus $\mu_m$ as $\lambda$ from zero to 40 when $\sin^2\theta_W=0.2312$. (d) The plot of magnetic charge $M$ in the U(1), SU(2) 't Hooft, electromagnetic and neutral magnetic field versus the compactified coordinate $\bar{x}$ when $\sin^2\theta_W=0.2312$ and $\lambda=1$.} 
\label{fig.2}
\end{figure}

Figure \ref{fig.2} (a) and (b) show the plots of total energy $E$ and magnetic dipole moment $\mu_m$ versus $\lambda^{1/2}$ when $\sin^2\theta_W=0.2312$. The total energy $E$ increases logarithmically whereas the magnetic dipole moment $\mu_m$ decreases exponentially fast with increasing $\lambda^{1/2}$. The graph of energy $E$ versus magnetic dipole moment $\mu_m$ as $\lambda$ varies from 0 to 40 and $\sin^2\theta_W=0.2312$ is shown in Figure \ref{fig.2} (c) and it is a non-increasing graph.

The graphs of magnetic charge $M$ versus the compactified coordinate $\bar{x}$ when $\lambda=1$ and $\sin^2\theta_W=0.2312$ are plotted in Figure \ref{fig.2} (d) for the U(1) magnetic field, SU(2) 't Hooft magnetic field, the electromagnetic field, and the neutral ${\cal Z}^0$ magnetic field. As expected there is zero net magnetic charge in the neutral ${\cal Z}^0$ field, however the net magnetic charge for the electromagnetic field is $\frac{2\pi}{e}$ which is one half of the Cho-Maison magnetic monopole charge. The fact that there is zero magnetic charge at $r=0$ that increases to one half of $\frac{4\pi}{e}$ at finite distance from the origin at $\bar{x}=0.9171$ or $r\approx 11$ to infinity ($\bar{x}=1$) shows that the magnetic charge is a finite length line charge.                        

\begin{figure}[tbh]
\centering
\hskip-0.1in
\includegraphics[width=6.0in,height=6.0in]{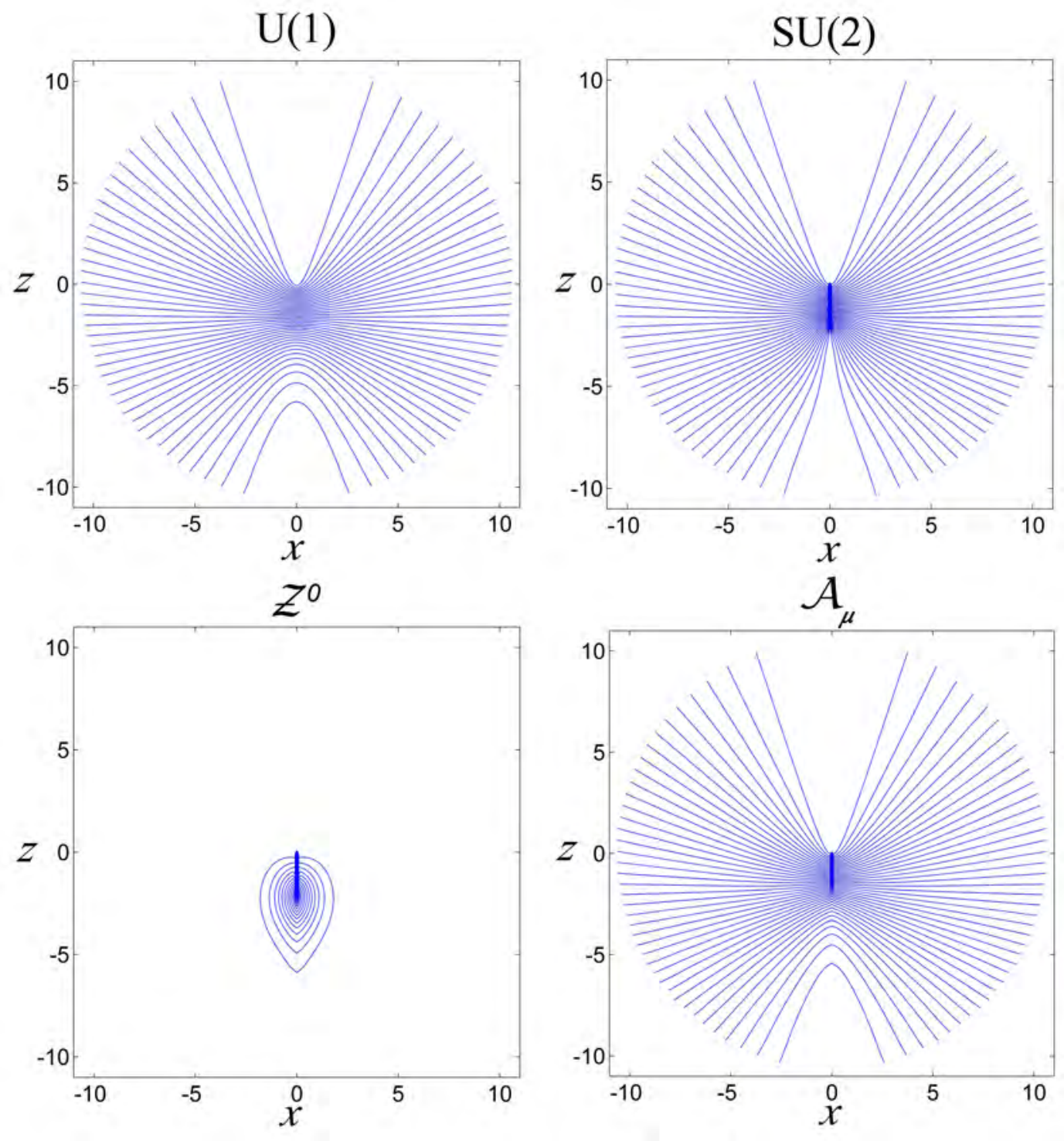} 
\caption{Contour line plots of the magnetic field lines of (a) the U(1) field, (b) the SU(2) 't Hooft field, (c) the neutral ${\cal Z}^0$ field, and (d) the electromagnetic field when $\sin^2\theta_W=0.2312$ and $\lambda=1$.}
\label{fig.3}
\end{figure}

With the U(1) magnetic field and the SU(2) 't Hooft magnetic field given by \cite{kn:18}
\begin{eqnarray}
g^\prime B_i^{U(1)} &=& -\epsilon^{ijk}\partial_j {\cal B}_S\partial_k\phi ~~~~\mbox{and}\nonumber\\
gB^{tHooft}_i &=& -\epsilon^{ijk}\partial_j\{gA^{\prime 3}_k\} \nonumber\\
&=& -\epsilon^{ijk}\partial_j \{(P_1 h_2-P_2 h_1) - (1-\cos\alpha)\}\partial_k\phi,
\label{eq.23}
\end{eqnarray}
and the definition of the electromagnetic and neutral ${\cal Z}^0$ field gauge potential (\ref{eq.14}), the magnetic field lines of the U(1) field, the SU(2) 't Hooft field, the neutral ${\cal Z}^0$ field, and the electromagnetic field can be drawn as shown in Figure \ref{fig.3} when $\lambda=1$ and $\sin^2\theta_W=0.2312$. 
The magnetic field lines of the half-monopole in the U(1) field clearly shows that the half-monopole is a one dimensional finite length line charge extending from the origin $r=0$ along the negative $z$-axis. Unlike the Cho-Maison one monopole there is no magnetic charge at the origin $r=0$.  The 't Hooft magnetic field lines pattern resembles that of the half-monopole in the SU(2) Georgi-Glashow model \cite{kn:9}. The difference between the 't Hooft magnetic field lines compare to that of the U(1) magnetic field lines is that the 't Hooft  magnetic field lines originate from a small volume centered at the origin $r=0$ and the lines run along a finite length of the negative $z$-axis before spreading out like hedgehog. In the U(1) magnetic field, the field lines originate from a finite length of the negative $z$-axis starting from $r=0$ to $r \approx 11$.


\begin{table}[tbh]
\begin{center}
\begin{tabular}{ccccccccccccc}
\hline
$\theta_W$ &	$10^o$	& $15^o$ &	$20^o$ &	$30^o$ &	$40^o$	& $45^o$ &	$50^o$ &	$60^o$ &	$70^o$ &	$80^o$ & $90^o$\\ \hline
$E$ &	0.530 &	0.565 &	0.590 &	0.629 &	0.657 &	0.667 &	0.675 &	0.684 &	0.685 &	0.682 &	0.681  \\ \hline
$\mu_m$	& 1.070 &	0.980 &	0.942 &	0.891 &	0.851 &	0.835 &	0.823 &	0.809 &	0.807 &	0.810 &	0.813 \\ \hline
\end{tabular}
\end{center}
\caption{Values of total energy $E$ in units of $\frac{1}{4\pi}$ and magnetic dipole $\mu_m$ of the one-half monopole for various values of Weinberg angle $\theta_W$ when $\lambda = \zeta = 1$.}
\label{table.1}
\end{table}

\begin{table}[tbh]
\begin{center}

\begin{tabular}{ccccccccccccc}
\hline
$\lambda$ &	0	& 0.1 &	0.5 &	1 &	2	& 4 &	8 &	10 &	20 & 30 & 40 \\ \hline
$E$ &	0.563 &	0.590 &	0.612 &	0.625 &	0.639 &	0.656 &	0.674 &	0.680 &	0.700 &	0.711 &	0.720  \\ \hline
$\mu_m$	& 1.028 &	0.958 &	0.916 &	0.897 &	0.877 &	0.858 &	0.840 &	0.834 &	0.816 &	0.806 &	0.799 \\ \hline
\end{tabular}
\end{center}
\caption{Values of total energy $E$ in units of $\frac{1}{4\pi}$ and magnetic dipole moment $\mu_m$ of the one-half monopole for various values of $\lambda$ when $\theta_W = 28.74^o$ and $\zeta = 1$.}
\label{table.2}
\end{table}

\section{Comments}
\label{sect.5}
The half-monopole in the Weinberg-Salam model is a one-half Cho-Maison monopole of magnetic charge $\frac{2\pi}{e}$. It can exist individually as a finite length line magnetic charge as presented in section \ref{sect.4} or as monopole-antimonopole pairs (MAP) with a ${\cal Z}^0$ field flux string joining the monopole and antimonopole in the sphaleron and sphaleron-antisphaleron pair \cite{kn:11}-\cite{kn:15}, \cite{kn:18}. In the sphaleron solutions, there is an electromagnetic current loop circulating each monopole-antimonopole pair. This half-monopole which is a line magnetic charge is different from the half-monopole in the Georgi-Glashow model where it is a point charge located at the origin \cite{kn:9} but both half-monopoles possess finite energy. 

A Cho-Maison monopole \cite{kn:4} and Cho-Maison monopole-antimonopole chains (MAC) \cite{kn:18} possess infinite energy and vanishing magnetic dipole moment but a half Cho-Maison monopole possesses finite energy and nonvanishing magnetic dipole moment whether they exist individually or in pairs in the sphaleron solutions. 

Further study of this half-monopole solution is carried out by introducing electric charge into the solution to create a half-dyon solution of the Weinberg-Salam theory and this work will be presented in a separate work.

\section*{Acknowledgements}

The authors would like to thank the Ministry of Science, Technology and Innovation (MOSTI) for the ScienceFund grant (305/PFIZIK/613613). 


\end{document}